\documentclass[showpacs, pra,onecolumn,preprintnumbers ,amsmath, amssymb, superscriptaddress, aps]{revtex4-2}
\usepackage{color}
\usepackage{amsmath,amssymb}
\usepackage{pifont}% http://ctan.org/pkg/pifont
\usepackage{amssymb}  % More symbols
\usepackage{bbold}
\usepackage{float}
\usepackage{subfloat}
\usepackage[squaren]{SIunits}

\usepackage[caption=false]{subfig}
\usepackage{tikz}
\usepackage{makecell}
\usepackage{subfig}
\usepackage{pifont}   % Ding symbols
\usepackage{graphicx} % Include figure files
\graphicspath{{Figures/}}
\usepackage{dcolumn}  % Align table columns on decimal point
\usepackage{bm}       % bold math
\usepackage{multirow} % Table functions
\usepackage{placeins}% For making floats not move around everywhere
\usepackage[colorlinks]{hyperref}
\usepackage{mathtools}
\usepackage{appendix}

\newcommand{\bb} {\color{blue}}

\begin{document}
	
	\title{Transport properties in gapped  graphene through magnetic barrier {in a} laser field }
	
	\author{Rachid El Aitouni}
	\affiliation{Laboratory of Theoretical Physics, Faculty of Sciences, Choua\"ib Doukkali University, PO Box 20, 24000 El Jadida, Morocco}
	
	\author{Miloud Mekkaoui}
	\affiliation{Laboratory of Theoretical Physics, Faculty of Sciences, Choua\"ib Doukkali University, PO Box 20, 24000 El Jadida, Morocco}
	
	\author{Ahmed Jellal}
	\email{a.jellal@ucd.ac.ma}
	\affiliation{Laboratory of Theoretical Physics, Faculty of Sciences, Choua\"ib Doukkali University, PO Box 20, 24000 El Jadida, Morocco}
	\affiliation{Canadian Quantum  Research Center,
		204-3002 32 Ave Vernon,  BC V1T 2L7,  Canada}
		\author{Michael Schreiber}
	\affiliation{Institut für Physik, Technische Universität, D-09107 Chemnitz, Germany}

	\begin{abstract}
We study the transport properties of Dirac fermions through gapped graphene through a magnetic barrier irradiated by a laser field oscillating in time. We use Floquet theory and the solution of Weber's differential equation to determine the energy spectrum corresponding to the three regions composing the system. The boundary conditions and the transfer matrix approach {are} employed to explicitly determine the transmission probabilities for multi-energy bands and the associated  conductance.  As an illustration, we focus only on the three first bands: the central band $T_0$ (zero photon exchange) and the two first side bands $T_{\pm1}$ (photon emission or absorption). It is found that  the laser field activates the process of translation through photon exchange. Furthermore, we show that varying the incident angle and energy gap strongly affects the transmission process. The conductance increases when the number of electrons that cross the barrier increases, namely when there is a significant transmission.

	\end{abstract}

		\pacs{78.67.Wj, 05.40.-a, 05.60.-k, 72.80.Vp\\
		{\sc Keywords}: Graphene, laser field, magnetic field, energy gap, transmission, Klein effect, conductance.}
	\maketitle

\section{Introduction}

Graphene is a two-dimensional carbon-based material that is one atom thick, and has atoms structured in a hexagonal shape like a honeycomb \cite{Novoselov2004,Novoselov2005}. {\bb Graphene possesses remarkable characteristics, including} a very high mobility  \cite{mobil2,mobil}, electrons moving with a speed 300 times lower than the {speed of light}, a good conductivity (minimal in the vicinity of the Dirac points, i.e., always the fermions pass), being flexible \cite{flix} and being very hard \cite{Beenakker2008}.
 Due to these properties, graphene is becoming the most used material in the technological industries \cite{Bhattacharjee2006,Bunch2005,Berger2004}. 
 It is theoretically studied in the framework of the tight-binding model \cite{Tight} and as a result, the energy spectrum shows a linear   dispersion relation. In addition,  the energy bands are in contact at six points \cite{Castro2009, propr}, called Dirac points K (K'), and form cones around them. It is surprising that electrons  can pass from the valance band to the conduction band easily without any effect. This lack of  {excitation} energy constitutes, in fact, an obstacle and a challenge for the fabrication of devices based on graphene. Consequently, to control the passage of electrons, an energy gap should be created between the two bands. Several studies have been reported on the subject to overcome such situations, for instance, either by deforming graphene to generate  pseudo-magnetic fields that play the role of a real magnetic field \cite{def1,def4} or by stacking one layer of graphene on the other \cite{Morozov2005,scatring}.

On the other hand, fermions confined in graphene under barriers, at normal incidence,  can cross them even if their energy is less than the barrier heights, an effect known as the Klein paradox \cite{klien2}. 
For an oscillating potential over time, the energy spectrum acquires sub-bands, generating several transmission modes, and each mode corresponds to an energy band \cite{jellal2014}.
Furthermore, an applied magnetic field to graphene generates a quantized energy spectrum known as Landau levels \cite{Landau,Magnetic2011,confinementmagnetic,conmagnetic}. Combining these with the oscillating potential gives rise to a current density in $x$- and $y$-directions \cite{Elaitouni2022}. When the graphene is irradiated by a time-varying laser field,  
subbands {emerge} in the energy spectrum, and then the barrier {exchanges} photons with the fermions, generating infinite transmission modes \cite{biswas2013,biswas2012,rachid2022,laser2,jellal2014}. As a consequence, the laser field suppresses the Klein effect, which makes it possible to control the passage of fermions.

We investigate how Dirac fermions can cross a gapped graphene subjected to a magnetic barrier and irradiated by a laser field. Within the framework of Floquet theory \cite{floquetappr} and by using the solution of Weber's differential equation \cite{grad}, we will be able to determine the eigenspinors corresponding to each region composing the system. These will be matched at boundaries and mapped in matrix form by applying the matrix transfer approach to finally get the transmission coefficients for all energy bands. Now, with the help of the current density, we derive the transmission probabilities for all modes. 
The conductance is also calculated by integrating the total transmission over all incident angles.
Since it is not easy to treat all modes numerically, we limit our study to the {first three} bands, which are the central band ($l=0$) and the two first side bands ($l =\pm1$). We show that increasing the barrier width, or the incidence energy, decreases the transmissions, which implies that the number of electrons that cross the barrier decreases, {consequently}, the conductance decreases. On the other hand, when the intensity of the laser field increases, we observe that the transmissions decrease {\bb but increase as long as the laser field frequency increases}. {When the barrier width increases,} it is found that the resonance peaks appear, and their number increases. Another set of results shows that the transmissions are almost {zero} when the incidence energy is less than the energy gap, and the Klein paradox is still present.

This paper is organized as follows. In Sec. \ref{TTMM}, we present the Hamiltonian describing our system and {\bb we  solve} the eigenvalue equations to determine the wave functions in the three regions. We use the boundary conditions and the matrix formalism to express the transmission probabilities of each band, and we calculate the integral of this total transmission which makes it possible to determine the conductance at zero temperature in Sec. \ref{TTCC}. We discuss our numerical results in Sec. \ref{NNRR}. Finally, we conclude our work.

\section{Theoretical model}\label{TTMM}

\begin{figure}[H]
	\centering
	\includegraphics[scale=0.5]{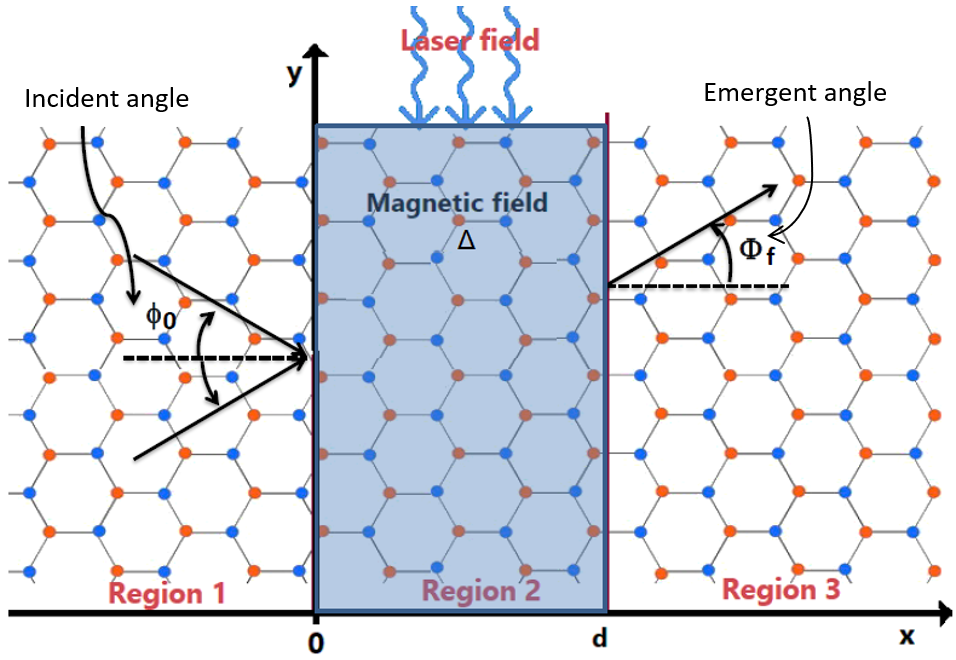}
	\caption{Schematic presentation of a graphene sheet irradiated by a linearly polarized monochromatic {laser feild} over a finite region of width $d$ subjected to a mass term {$\Delta$} and magnetic field.}
	\label{fig1}
\end{figure}

We study the behavior of Dirac fermions in {a} graphene sheet divided into three regions. Regions 1 and 3 contain only pristine graphene, whereas the gapped region 2 of width $d$ is subjected to  a perpendicular  magnetic field and irradiated by a laser field, as shown in  Fig. \ref{fig1}. 

The present system can be described the following Hamiltonian
\begin{equation}
	H= v_F \vec\sigma\cdot \left[\vec p-\frac{e}{c}\left({\vec A_L(t)+\vec A_B(x)}\right)\right]+\Delta \sigma_z
\end{equation}
where $\sigma_{x,y,z}$ are Pauli matrices, $v_F\approx c/300$ is the {Fermi velocity} , $\vec{p}=-i\hbar\left(\frac{\partial}{\partial x},\frac{\partial}{\partial y}\right)$ the momentum operator, $e$ the electronic
charge. The vector potential
 $\vec{A}_L(t)$ of the laser field in the dipole approximation \cite{approx} is generated by an electric field of amplitude $F$ and frequency $\omega$ defined as $E(t)=F\sin(\omega t)$, which is given by
\begin{equation}
	{\vec{A}_L(x,y,t)}=(0,A_0\cos(\omega t),0)
\end{equation}
{with the laser field amplitude} $A_0=\frac{F}{\omega}$. {For the magnetic field,} the vector potential $\vec{A}_B( x)$ is chosen in the Landau gauge $B(0,x,0)$ and the continuity allows us to write
\begin{equation}
	{\vec A_B(x)}=	\left \{
	\begin{array}{ll}
		0, &x<0 \\
	Bx,  & 0<x<d\\
		Bd,  &  x>d.
	\end{array}
	\right.
\end{equation}

To determine the eigenspinors $    \Psi(x,y,t)=(\Psi_1, \Psi_2)^T$ in the three regions, we solve the eigenvalue equation, with $T$ standing for transpose. In region 2 ($0<x<d$), we get 
\begin{align}
&	\Delta \Psi_1(x,y,t) + v_F\left[p_x-i\left(p_y-\frac{eF}{\omega}\cos(\omega t)-eBx\right)\right]\Psi_2(x,y,t)=i\hbar\frac{\partial }{\partial t}\Psi_1(x,y,t)\label{val1}\\
&	v_F\left[p_x+i\left(p_y-\frac{eF}{\omega}\cos(\omega t)-eBx\right)\right]\Psi_1(x,y,t)-\Delta \Psi_2(x,y,t)=i\hbar\frac{\partial }{\partial t}\Psi_2(x,y,t)\label{val2}
\end{align}
To proceed further, note that in the framework of the Floquet approximation \cite{floquetappr}, the oscillation of the laser field over time produces several energy modes in the eigenspinors. As a result, we have %have the total wave function takes the following form.
%	\begin{equation}
		$\Psi(x,y,t)=\psi(x,y,t)e^{-\frac{iEt}{\hbar}}$
%	\end{equation}
where $E$ is the Floquet quasi-energy, $\psi(x,y,t)$ is a time periodic function satisfying $\psi(x,y,t+t_0)=\psi(x,y,t)$ and $t_0$ is the time period of the laser field. On the other hand, if the Hamiltonian is invariant along the $y$-direction, then we write  $\Psi(x,y,t)=e^{ik_yy}e^{-\frac{iEt}{\hbar}}\varphi(t)(\phi_1(x),\phi_2(x))^T$, and therefore (\ref{val1},\ref{val2}) become
\begin{eqnarray}
	v_F\left[-i\frac{\partial}{\partial x}-i\left(k_y-\frac{F}{\omega}\cos(\omega t)-Bx\right)\right]\phi_2(x)\varphi(t)e^{ik_yy}e^{-iEt}&=&\left(i\frac{\partial }{\partial t}-\Delta\right)\phi_1(x)\varphi(t)e^{ik_yy}e^{-iEt}\label{d1}\\
v_F\left[-i\frac{\partial}{\partial x}+i\left(k_y-\frac{F}{\omega}\cos(\omega t)-Bx\right)\right]\phi_1(x)\varphi(t)e^{ik_yy}e^{-iEt}&=&
\left(i\frac{\partial }{\partial t}+\Delta\right)\phi_2(x)\varphi(t)e^{ik_yy}e^{-iEt}\label{d2}
\end{eqnarray}
in the system unit ($\hbar=e=c=1$). It is straightforward to find 
$-i\frac{F}{\omega}\cos(\omega t)=\frac{\partial }{\partial t}\varphi(t)$
and therefore the temporal component is 
\begin{eqnarray}\label{tm1}
\varphi(t)=e^{-i\alpha \sin(\omega t)}.
\end{eqnarray}
Now, {we use} the Jacobi–Anger identity $e^{-i\alpha\sin(\omega t)}=\sum_{-\infty}^{+\infty}J_m(\alpha)e^{-im\omega t}$ to write (\ref{d1},\ref{d2}) as
\begin{eqnarray}
\frac{\partial \phi_2(x)}{\partial x}-\left[\frac{x}{\ell_B^2}-k_y+m\varpi\right]\phi_2(x)-i (\varepsilon+m\varpi-\delta)\phi_1(x)=0\label{E10}\\
\frac{\partial \phi_1(x)}{\partial x}+\left[\frac{x}{\ell_B^2}-k_y+m\varpi\right]\phi_1(x)-i (\varepsilon+m\varpi+\delta)\phi_2(x)=0\label{E11}
\end{eqnarray}
where $\ell_B=\frac{1}{\sqrt{B}}$, $\varpi=\frac{\omega}{v_F}$, $\tilde{F}=\frac{F}{v_F}$, $\varepsilon=\frac{E}{v_F}$ and $\delta=\frac{\Delta}{v_F}$. From
 (\ref{E10},\ref{E11}), we obtain two new decoupled equations 
\begin{eqnarray}
	\frac{\partial^2\phi_{1}(x)}{\partial ^2 x}+\left[\frac{1}{\ell_B^2}-\left(\frac{x}{\ell_B^2}-k_y+m\varpi\right)^2+(\varepsilon+m\varpi)^2-\delta^2\right]\phi_{1}(x)&=&0\label{P1}\\
	\frac{\partial^2\phi_{2}(x)}{\partial ^2 x}+\left[-\frac{1}{\ell_B^2}-\left(\frac{x}{\ell_B^2}-k_y+m\varpi\right)^2+(\varepsilon+m\varpi)^2+\delta^2\right]\phi_{1}(x)&=&0\label{P2}.
\end{eqnarray}
These can be expressed in terms of the Weber differential equations \cite{grad,math} by making the {change of variable} $X_m=\sqrt{2}\left(\frac{x}{\ell_B}-k_y\ell_B+m\varpi \ell_B\right)$ and setting $v_m=\frac{(\varepsilon \ell_B+m\varpi \ell_B)^2-(\delta \ell_B)^2}{2}$, to get
\begin{eqnarray}
	\frac{d^2\phi_{1,2}(X_m)}{dX_m^2}+\left[\pm\frac{1}{2}-\frac{X^2_m}{4} +v_m\right]\phi_{1,2}(X_m)=0\label{web}
\end{eqnarray}
having the following solutions 
%The solution of Eqs. (\ref{web}) can be written as \cite{grad,math}
\begin{eqnarray}
	\phi_1(X_m)&=&A_mD_{v_m}(X_m)+B_mD_{v_m}(-X_m)\\
		\phi_2(X_m)&=&-\frac{i\sqrt{2 }}{\varepsilon \ell_B+m\varpi \ell_B+\delta \ell_B}\left[ A_mD_{v_m+1}(X_m)-B_mD_{v_m+1}(-X_m)\right]
\end{eqnarray}
where $A_m, B_m$ are constant coefficients corresponding to $m$th side-band, and $D_{v_m}$ is the parabolic cylinder function. Consequently, the eigenspinors in  region $2$ take the form
\begin{equation}
	\Psi_{2}(x,y,t)=e^{ik_yy}\sum_{l=-\infty}^{+\infty}\left[A_{l}\begin{pmatrix}
		\Xi^+_l(x)\\\ \eta^+_l(x)
	\end{pmatrix}
	+B_{l}\begin{pmatrix}
		\Xi^-_l(x)\\ \eta^-_l(x)
	\end{pmatrix}\right]\sum_{m=-\infty}^{+\infty}J_{m}(\alpha)e^{-i\left(\varepsilon+(l+m)\omega\right)t}
\end{equation}
and we have defined
\begin{eqnarray}
	\Xi^\pm(x)&=&D_{v_m}\left(\pm X_m\right)\\
	\eta^\pm(x)&=&\mp\frac{i\sqrt{2}}{\varepsilon \ell_B+m \varpi \ell_B+\delta \ell_B} D_{v_m+1}\left(\pm X_m\right).
\end{eqnarray}

In the region 1 ($x<0$) we have only pristine graphene, and then we can easily obtain the associated eigenspinors and eigenvalues \cite{jellal2014}
\begin{align}
&\Psi_1(x,y,t)=e^{ik_yy}\sum_{m=-\infty}^{+\infty} \left[\delta_{l,0}\begin{pmatrix}
		1\\ \Lambda_l
	\end{pmatrix}e^{ik_lx}+\sum_{m,l=-\infty}^{+\infty}r_l\begin{pmatrix}
		1\\-\Lambda^*_l
	\end{pmatrix}e^{-ik_lx}\right]\delta_{m,l}e^{-iv_F( \varepsilon+m\varpi)t}
\\
&	\varepsilon+l\varpi=s_l\sqrt{k^2_l+k^2_y}
\end{align}
where $r_l$ is the amplitude of the reflected wave corresponding to band $l$, $\delta_{m,l}=J_{m-l}(\alpha=0)$, $	s_l=\text{sgn}(v_F\varepsilon+lv_F\varpi)$,
$	\phi_l=\tan^{-1}\frac{k_y}{k_l}$,
$	k_l=\varepsilon\cos{\phi_l}$, 
$	k_y=\varepsilon\sin{\phi_l}$ and
\begin{align}
\Lambda_l=s_l\frac{k_l+ik_y}{\sqrt{k^2_l+k^2_y}}=s_le^{i\phi_l}.
\end{align}
 We can establish
the relation between the incident angles 
\begin{equation}
	\phi_l=\arcsin\left(\frac{\varepsilon}{\varepsilon+l\varpi}\sin(\phi_0)\right).
\end{equation}

In region 3 ($x>d$), the emergent angle $\phi'_l$ is different than the incident one $\phi_0$ because of the continuity of the vector potential. The solution is  \cite{jellal2014}
\begin{align}
&	\Psi_{3}(x,y,t)=e^{ik_yy}\sum_{m,l=-\infty}^{+\infty}\left[t_l\begin{pmatrix}
		1\\ \Lambda'_l
	\end{pmatrix}e^{ik'_lx}+b_l\begin{pmatrix}
		1\\-\Lambda'^*_l
	\end{pmatrix}e^{-ik'_lx}\right]\delta_{m,l}e^{-iv_F(\varepsilon+m\varpi)t}\\
&	\varepsilon+l\varpi =s_l\sqrt{k_l^{'2}+\left(k_y-\frac{ d}{\ell_B^2}\right)^2}
\end{align}
where $t_l$ is the transmission amplitude of the transmitted wave corresponding to the band $l$, ${b_l}$  is a null vector,
$\phi'_l=\tan^{-1}\frac{ky-\frac{ d}{\ell_B^2}}{k'_l}$, 
$k'_l=(\varepsilon+l\varpi)\cos{\phi'_l}$, 
$k_y=(\varepsilon+l\varpi)\sin{\phi'_l}+\frac{d}{\ell_B^2}$
and
\begin{align}
	\Lambda'_l=s_l\frac{k'_l+i\left(k_y-\frac{d}{\ell_B^2}\right)}{\sqrt{k_l^{'2}+\left(k_y-\frac{d}{\ell_B^2}\right)^2}}=s_le^{i\phi'_l}.
\end{align}

From the conservation of the momentum $k_y$, we get  the relation
\begin{equation}
	\phi'_l=\arcsin{\left(\frac{\varepsilon}{\varepsilon+l \varpi }\sin{\phi_0}-\frac{\frac{ d}{\ell_B^2}}{\varepsilon+l\varpi}\right)}.
\end{equation}
As we will see, the above results can be used to study the transport properties of gapped graphene scattered by a magnetic barrier and irradiated by a laser field. {We obtain} the transmissions associated with several energy bands and the corresponding conductance.

\section{Transmission probabilities}\label{TTCC}

We use the continuity of the eigenspinors {at} $x=0$ and $x =d$ to 
determine the transmission probabilities for the present system. This corresponds to the processes   
 $\Psi_1(0,y,t)=\Psi_{2}(0,y,t)$ and $\Psi_{2}(d,y,t)=\Psi_{3}(d,y,t)$,
 which yields 
\begin{align}
&	\delta_{m,0}+r_m=\sum_{l=-\infty}^{+\infty}\left(A_l\Xi^+_l(0)+B_l\Xi^-_l(0)\right)J_{m-l}(\alpha)\\
&	\delta_{m,0}\Lambda_m-r_m\Lambda_m^*=\sum_{l=-\infty}^{+\infty}\left(A_l\eta^+_l(0)+B_l\eta^-_l(0)\right)J_{m-l}(\alpha)\\
&	t_me^{ik'_md}+b_me^{-ik'_md}=\sum_{l=-\infty}^{+\infty}\left(A_l\Xi^+_l(d)+B_l\Xi^-_l(d)\right)J_{m-l}(\alpha)\\
&	t_m\Lambda^{'}_me^{ik'_md}-b_m\Lambda_m^{'*}e^{-ik'_md}=\sum_{l=-\infty}^{+\infty}\left(A_l\eta^+_l(d)+B_l\eta^-_l(d)\right)J_{m-l}(\alpha).
\end{align}
We have four equations, but each one has an infinite number of modes, and to solve the problem, we use the transfer matrix approach. As a result, we get 
\begin{equation}\label{35}
	\begin{pmatrix}
		\Upsilon_1\\
		\Upsilon'_1
	\end{pmatrix}
	=\begin{pmatrix}
		\mathbb{N}_{1,1}&	\mathbb{N}_{1,2}\\
		\mathbb{N}_{2,1}&	\mathbb{N}_{2,2}
	\end{pmatrix}	\begin{pmatrix}
		\Upsilon_2\\
		\Upsilon'_2
	\end{pmatrix}=\mathbb{N}
	\begin{pmatrix}
	\Upsilon_2\\
	\Upsilon'_2
\end{pmatrix}
\end{equation}
with
\begin{equation}
	\mathbb{N}=\begin{pmatrix}
		\mathbb{I} & \mathbb{I} \\
		\mathbb{\Gamma}^+&\mathbb{\Gamma}^-\\
	\end{pmatrix}^{-1}
	\begin{pmatrix}
		\mathbb{X}^+_0&\mathbb{X}^-_0\\
		\mathbb{R}^+_0&\mathbb{R}^-_0
	\end{pmatrix}
	\begin{pmatrix}
	\mathbb{X}^+_d&\mathbb{X}^-_d\\
	\mathbb{R}^+_d&\mathbb{R}^-_d
\end{pmatrix}^{-1}
	\begin{pmatrix}
		\mathbb{I} & \mathbb{I} \\
		\mathbb{\Gamma}'^+&\mathbb{\Gamma}'^-\\
	\end{pmatrix}
	\begin{pmatrix}
		\mathbb{K}^+&\mathbb{O} \\
		\mathbb{O}&\mathbb{K}^-\\
	\end{pmatrix}
\end{equation}
and
\begin{eqnarray}
	\mathbb{\Gamma}^{\pm}=\pm\delta_{m,l}\Lambda_l^{\pm1}, \quad	\mathbb{\Gamma'}^{\pm}=\pm\delta_{m,l}\Lambda_l^{'\pm1}, \quad
	\mathbb{X}^\pm_z=\Xi_l^\pm(z)J_{m-l}(\alpha),\quad
	\mathbb{R}^\pm_z=\eta_l^\pm(z)J_{m-l}(\alpha),\quad
	\mathbb{K}^\pm=e^{\pm ik'_lL}\delta_{m,l}
\end{eqnarray}
where $\mathbb{O}$ is the zero matrix, $\mathbb{I}$ is the unit matrix and $z=\{0,d\}$. 
In this case, we take into account Dirac fermions traveling from left to right with energy $E$, and from \eqref{35}, we obtain
\begin{equation}
	\Upsilon_2=\mathbb{N}^{-1}_{1,1}\Upsilon_1
\end{equation}
with  the Kronecker coefficient  $\delta_{0,l}=\Upsilon_1$ and $\Upsilon_2=t_l$.

Because {$n$ and $l$ } range from $-\infty $ to $+\infty $ and are challenging to solve, the aforementioned transfer matrix is of infinite order. Due to this, we replace the infinite series with a finite set of terms ranging from $-N$ to $N$, provided that $N\ge \frac{F}{\omega^2}$ \cite{jellal2014},  resulting in 
\begin{equation}
	t_{-N+k}=\mathbb{N}'[k+1,N+1]
\end{equation}
where $\mathbb{N}'=\mathbb{N}^{-1}_{11}$, $k=0, 1, 2,\cdots N$.
To simplify, we limit our studies only to the {central band and the first two} side bands $l=0,\pm 1$ of energy $E\pm h\omega$ having the following transmission coefficients
\begin{eqnarray}
	t_{-1}=\mathbb{N}'[1,2], \quad
	t_{0}=\mathbb{N}'[2,2], \quad
	t_{1}=\mathbb{N}'[3,2].
\end{eqnarray}

On the other hand, the current density is determined from the {continuity equation}, its expression given by $	J=e v_F \Psi^{\mathbb{*}}\sigma_x\Psi$ , therefore the expression of the incident, reflected and transmitted current density given by
\begin{align}
&	J_{\text{inc},0}=ev_F(\Lambda_0+\Lambda^*_0)\\
&	J_{\text{tra},l}=ev_Ft^*_lt_l(\Lambda'_l+\Lambda'^*_l)\\
&	J_{\text{ref},l}=ev_Fr^*_lr_l(\Lambda_l+\Lambda^*_l)
\end{align}
The relation between the current density and the transmission probability is expressed as $	T_l=\frac{J_{\text{tra},l}}{J_{\text{inc},0}}$. Then, after some algebra, we get  
\begin{eqnarray}
T_{l}=\frac{\cos\phi'_{l}}{\cos\phi_0}|t_l|^2
\end{eqnarray}
and {the} total transmission probability is {given} by summing up over all modes
\begin{equation}
	T=\sum_{l}T_l.
\end{equation}

By definition, the conductance at zero temperature is the average of the flux of the fermions on the half Fermi surface \cite{conduct1, conduct2}, on the other hand it is the integration of the total transmission $T$ over $k_y$ \cite{Biswas2021}, given by
\begin{equation}
G=\frac{G_0}{2\pi}\int_{-k_y^{\text{max}}}^{k_y^{\text{max}}}T dk_y
\end{equation}
where $G_0$ is the conductance unit.
Using the relation between transverse wave vector $k_y$ and the incident angle $\phi_0$ to express $G$ as
\begin{equation}
	G=\frac{G_0}{2\pi}\int_{-\frac{\pi}{2}}^{\frac{\pi}{2}}T \cos\phi_0d\phi_0.
\end{equation}
To investigate and underline the basic features of the present system, we numerically analyze the transport properties based on the transmission channels and associated conductance {in the following chapter.}

\section{Results and discussion}\label{NNRR}

 \begin{figure}[ht]
	\centering
	\subfloat[]{\centering\includegraphics[scale=0.63]{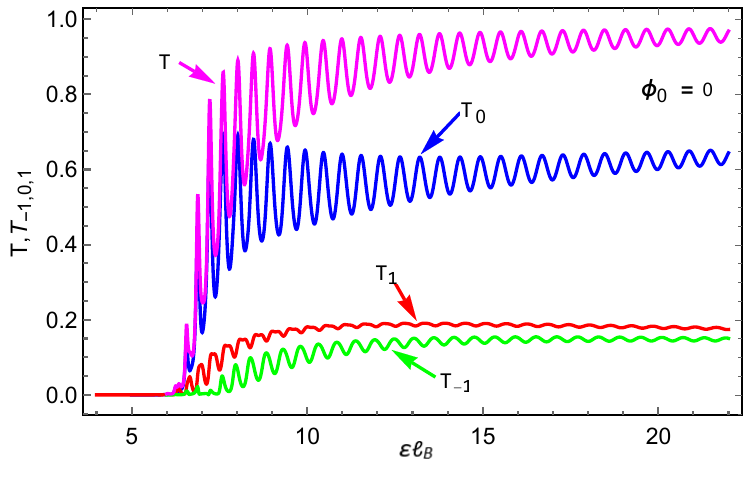}\label{fig2a}}\ \ \ \ \ \ \ \ \subfloat[]{\centering\includegraphics[scale=0.6]{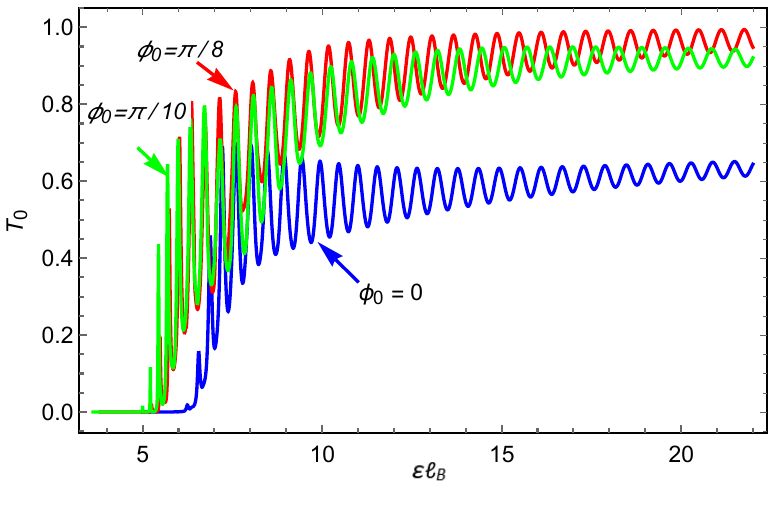}\label{fig2b}}\\
	\subfloat[]{\centering\includegraphics[scale=0.6]{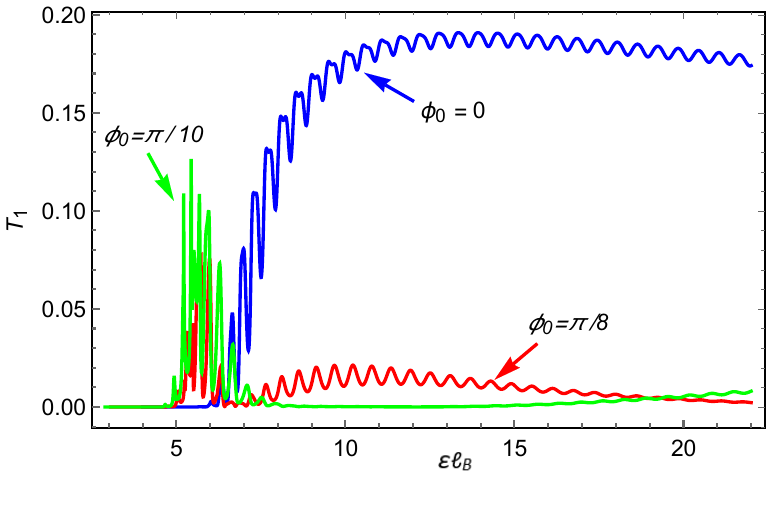}\label{fig2c}}\ \ \ \ \ \ \ \
	\subfloat[]{\centering\includegraphics[scale=0.6]{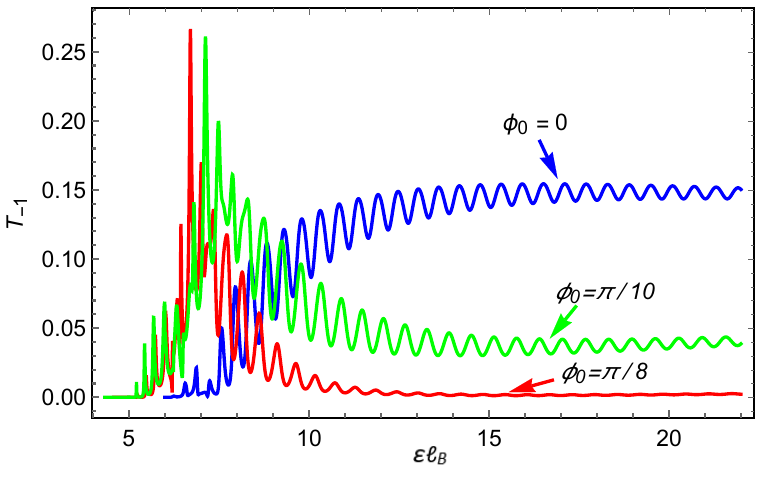}\label{fig2d}}
	\caption{{(Color online) Transmission probabilities as a function of the energy $\varepsilon \ell_B$ with $\tilde{F} \ell_B^2=0.6$, $\varpi \ell_B=1$, $\frac{d}{\ell_B}=5$ and $\delta \ell_B=5$, (a): $\phi_0=0$, $T$ (magenta line) total transmission probability,  $T_{0}$ (blue line), $T_{-1}$ (green line) and $T_{1}$ (red line), (b, c, d): $T_0, T_1, T_{-1}$ with $\phi_0=\{0, \frac{\pi}{10}, \frac{\pi}{8} \}$}.}\label{fig2}
\end{figure}

We numerically study  the transmission probabilities of Dirac fermions in gapped graphene through a  magnetic barrier {in} a laser field. Recall that the oscillation of the barrier over time generates several energy bands, which give rise to transmission channels. Due to the difficulty of analyzing all  modes, we will limit ourselves to the first three  bands, where the central band $T_0$ {corresponds to} zero photon exchange and the first two  side bands $T_{\pm1} $ {to} absorption or emission of photons.
Fig. \ref{fig2} shows the transmission probability as a function of the energy $\varepsilon \ell_B$ for different incident angles. There is transmission if the condition {$\varepsilon >\frac{\frac{d}{\ell_B^2}-l \varpi}{1+\sin{\phi_0}}$}
 is satisfied, in other words, this quantity plays the role of an effective mass \cite{Mekkoui2021}. For normal incidence, as depicted in Fig. \ref{fig2a}, transmission is {zero} for $\varepsilon<\delta$. {Due to} this condition, resonance peaks appear with decreasing amplitudes along the $\varepsilon \ell_B$-axis, {that is to say} the disappearance of the Fabry-P\'erot resonance, which is in agreement with previous results \cite{MEKKAOUI2018,Elaitouni2022}. The transmission process with zero photon exchange, $T_0$, is dominating, and therefore, the majority of the electrons cross the barrier without photon exchange.
{\bb Fig. \ref{fig2b} shows the behavior of $T_0$ for different incident angles, and as a result, we observe that $T_0$ increases sharply away from the normal incidence}. On the other hand, transmission with photon exchange {as shown in Figs. \ref{fig2c}, \ref{fig2d} there is a decrease for large energy}.
 We can conclude that the behavior of $T_0$ changes if we move away from the normal incidence and {that} 
  the photon exchange process {is suppressed}.

 \begin{figure}[ht]
 	\centering\subfloat[]{\centering\includegraphics[scale=0.67]{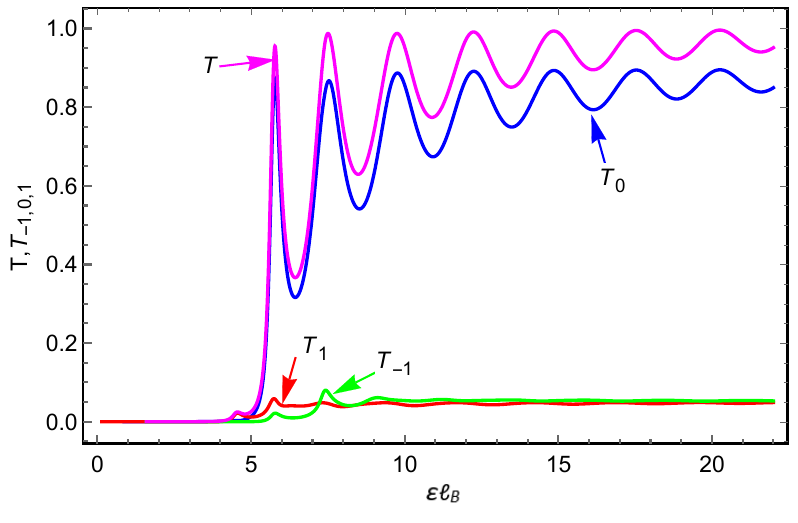}\label{fig3a}} \subfloat[]{\centering\includegraphics[scale=0.67]{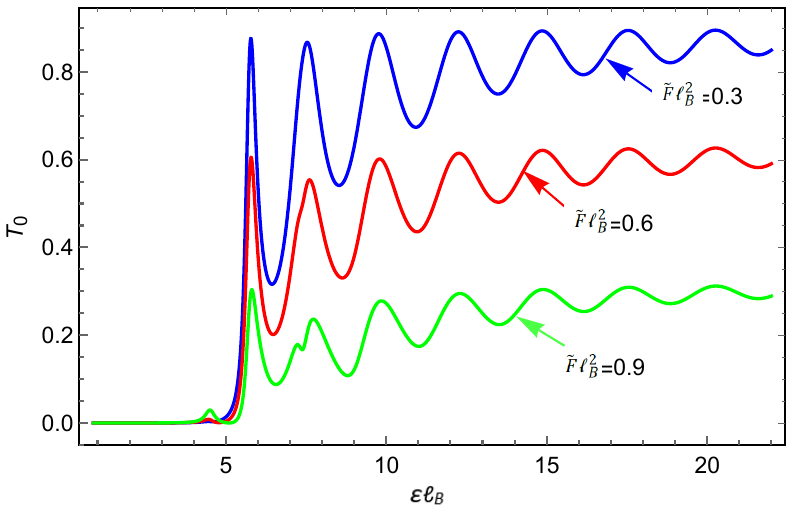}\label{fig3b}}\\
 	\subfloat[]{\centering\includegraphics[scale=0.67]{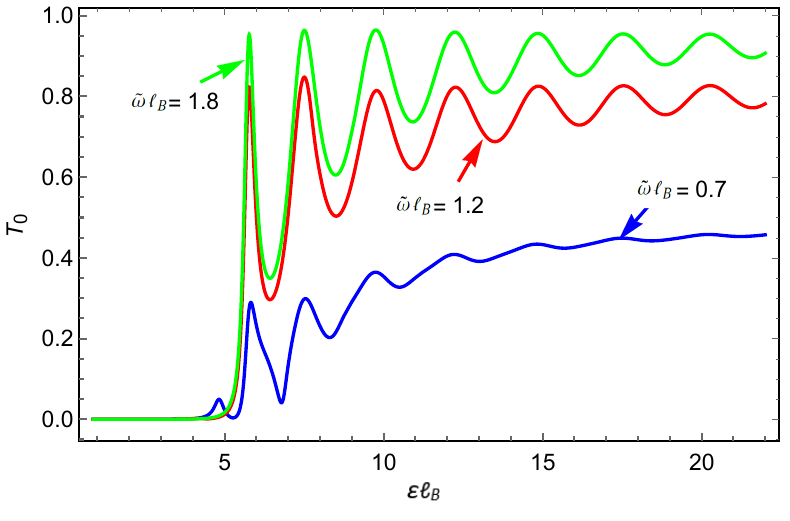}\label{fig3c}}
 	\subfloat[]{\centering\includegraphics[scale=0.67]{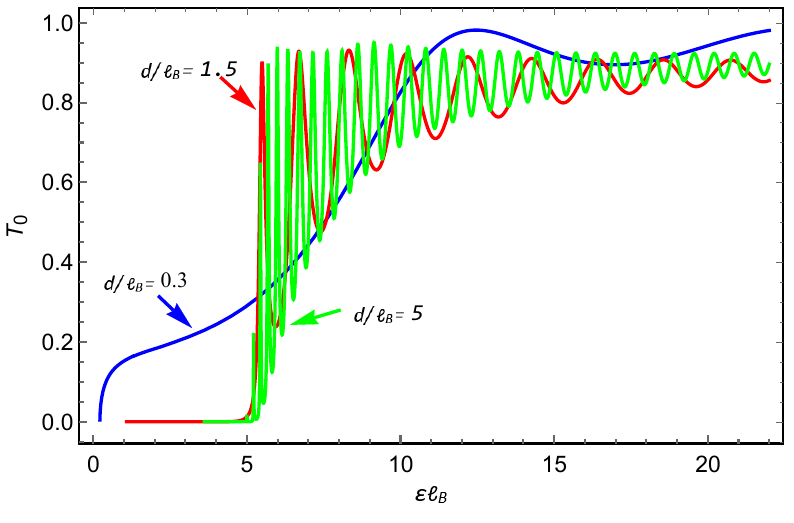}\label{fig3d}}
 	\caption{{(Color online) Transmission probabilities as a function of $\varepsilon \ell_B$ with $\delta \ell_B=5$, $\phi_{0}=\frac{\pi}{8}$ (a): $\tilde{F} \ell_B^2=0.3$, $\varpi \ell_B=1$, $\frac{d}{\ell_B}=1.2$, $T$ (magenta line), $T_{0}$ (blue line), $T_{-1}$ (green line) and $T_{1}$ (red line), (b): $\varpi \ell_B=1$, $\frac{d}{\ell_B}=1.2$, $\tilde{F} \ell_B^2=\{0.3, 0.6, 0.9\}$, (c):  $\frac{d}{\ell_B}=1.2$, $\tilde{F} \ell_B^2=0.5$, $\varpi \ell_B=\{0.7, 1.2, 1.8\}$, (d): $\tilde{F} \ell_B^2=0.5$, $\varpi \ell_B=1.5$, $\frac{d}{\ell_B}=\{0.3, 1.5, 5\}$.}}
 	\label{fig3}
 \end{figure}

Fig. \ref{fig3} displays the transmission probability as a function of $\varepsilon \ell_B$ under a suitable choice of physical  parameters. Transmissions appear when  condition $\varepsilon >\delta $ is satisfied. As clearly seen in Fig. \ref{fig3a}, we observe the dominance of $T_0$ compared to those corresponding to {the first two side bands}, and it is almost equal to the total transmission {as}
found
 in \cite{confinementmagnetic}. Now for different values of $\tilde{F} \ell_B^2$, we plot $T_0$ in Fig. \ref{fig3b}. We see that $T_0$ decreases with the increase of $\tilde {F} \ell_B^2$, because the increase in laser field suppresses $T_0$ as we have already seen \cite{rachid2022}. 
 Fig. \ref{fig3c} displays the effect of field frequency on transmission: increasing the frequency increases $T_0$. 
 Fig. \ref{fig3d} is drawn for different values of barrier width $\frac{d}{\ell_B}$. {If this increases,} resonance peaks appear and their number increases, and the
  oscillations get closer. A similar result is obtained in our previous work \cite{Makkoui2015}.

 \begin{figure}[ht]
 	\centering\subfloat[]{\centering\includegraphics[scale=0.67]{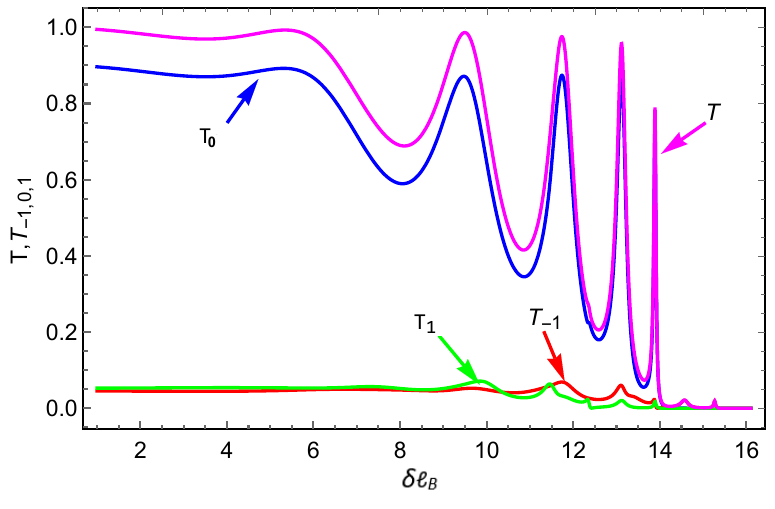}\label{fig4a}} \subfloat[]{\centering\includegraphics[scale=0.68]{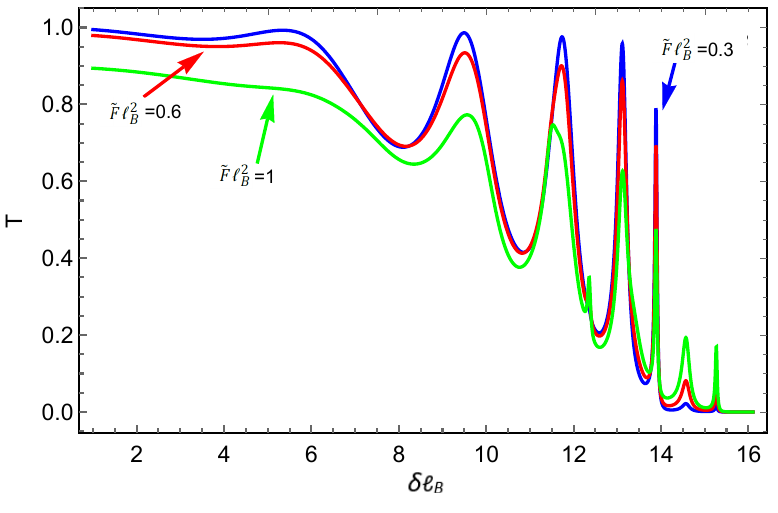}\label{fig4b}}\\
 	\subfloat[]{\centering\includegraphics[scale=0.66]{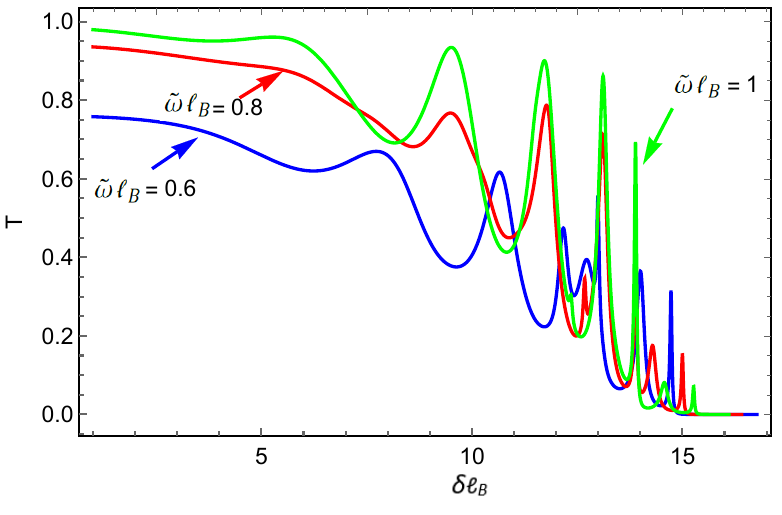}\label{fig4c}}
 	\subfloat[]{\centering\includegraphics[scale=0.66]{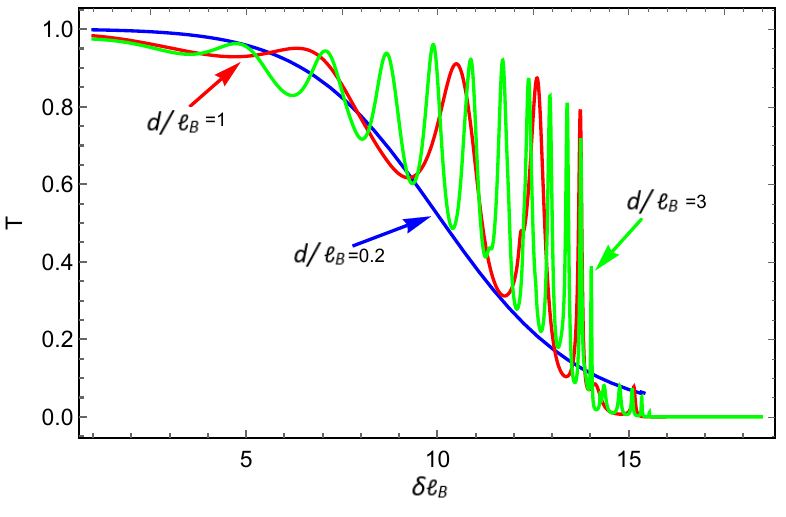}\label{fig4d}}
 	\caption{{(Color online) Transmission probabilities as a function of $\delta \ell_B$ with $\varepsilon \ell_B=15$ and $\phi_{0}=\frac{\pi}{8}$, (a): $\frac{d}{\ell_B}=1.2$, $\varpi \ell_B=1$ and $\tilde{F} \ell_B^2=0.3$, (b): $\frac{d}{\ell_B}=1.2$, $\varpi \ell_B=1$ and $\tilde{F} \ell_B^2=\{0.3, 0.6, 1\}$, (c): $\frac{d}{\ell_B}=1.2$, $\tilde{F} \ell_B^2=0.6$ and $\varpi \ell_B=\{0.6, 0.8, 1\}$}, (d): $\varpi \ell_B=1$, $\tilde{F} \ell_B^2=0.6$ and $\frac{d}{\ell_B}=\{0.2, 1, 3\}$.}
 	\label{fig4}
 \end{figure}

Fig. \ref{fig4} presents the transmission probabilities as a function of the energy gap $\delta \ell_B$. 
We show in Fig. \ref{fig4a} the total transmission probability (magenta line) and those with or without photon exchange.
We distinguish two interesting cases: first, for $\delta \ell_B<6$, the Klein effect is very clear and transmission with photon exchange is almost zero, that {means that} the majority of electrons cross the barrier without photon exchange. Second, for $\delta \ell_B > 6 $, the transmissions decrease in an oscillatory way until they become {zero} when $\delta \ell_B$ is close to $\varepsilon \ell_B=15$.
Fig. \ref{fig4b} displays the total transmission for different values of $\tilde{F} \ell_B$, and we see that the increase of $\tilde{F} \ell_B$ suppresses the transmission, as has been found in \cite{biswas2012}. The Klein effect is clear for very small values of $\tilde{F} \ell_B$ and $\delta \ell_B$. For $\tilde{F} \ell_B=0.3$,  the Klein effect is observed only for $\delta \ell_B<6$, then the transmission decreases in an oscillatory way until the {oscillations vanish}. If we increase $\tilde{F} \ell_B$ the transmission keeps the same shape with decreasing amplitude, which is in agreement with the results of \cite{MEKKAOUI2018}.
Fig. \ref{fig4c} is  similar to the previous one, but here we vary $\varpi \ell_B$. As a result,  for $\varpi \ell_B=1$ the Klein effect always exists up to $\varpi \ell_B=5$, then the transmission decreases in an oscillatory way towards zero near $\varepsilon \ell_B$. {On the other hand}, there will be total reflection if the incident energy is lower than the energy gap. 
If the frequency decreases, the transmission retains the same shape, but the amplitude decreases. Fig. \ref{fig4d} shows the effect of the barrier width on the total transmission. We observe that resonance peaks appear when the width increases. For very small widths, the Klein effect {is found} up to $\delta \ell_B$ {$\approx$} $6$, and then the transmission decreases towards zero. Increasing the width increases the number of oscillations and their amplitudes, as already seen in \cite{Makkoui2015}. We summarize that  increasing the amplitude of the field suppresses transmission inside the barrier. On the other hand, increasing the frequency increases the transmission, and increasing the width increases the number of oscillations and their amplitude.

\begin{figure}[ht]
	\centering
\subfloat[]{\centering\includegraphics[scale=0.68]{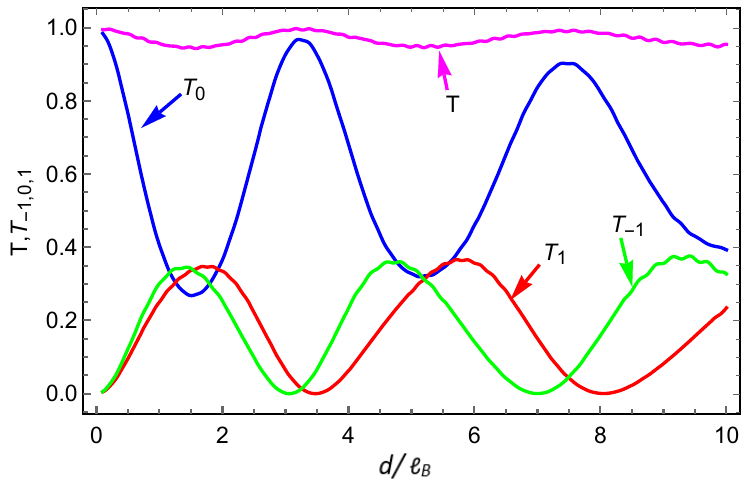}\label{fig5a}} \subfloat[]{\centering\includegraphics[scale=0.69]{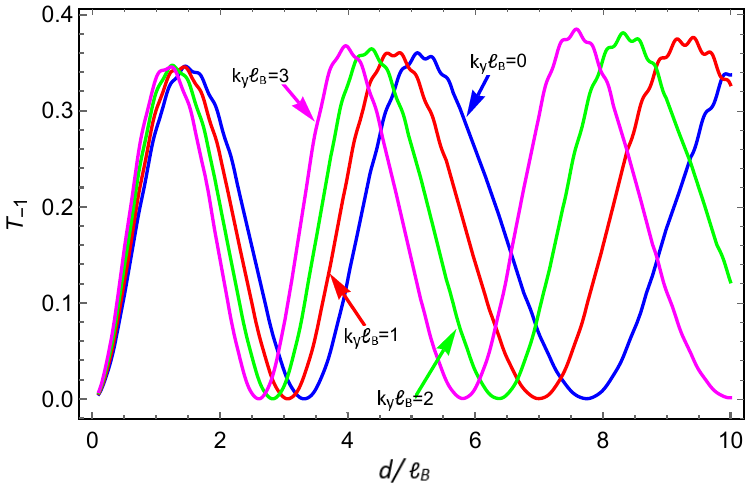}\label{fig5b}}\\
\subfloat[]{\centering\includegraphics[scale=0.69]{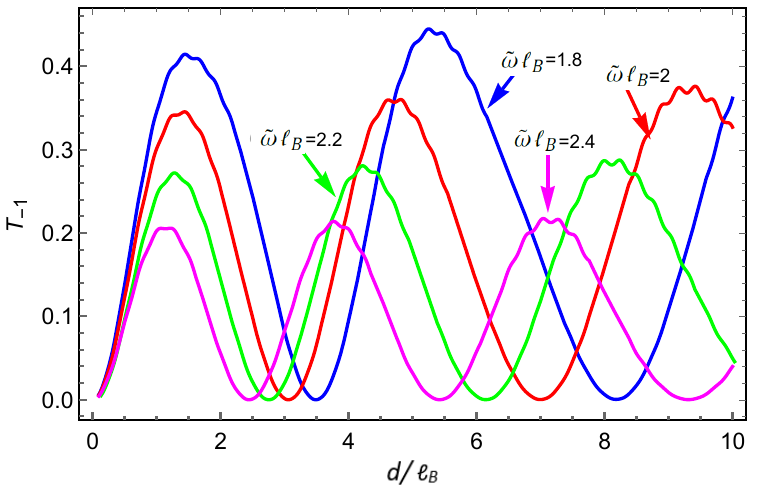}\label{fig5c}} 
\subfloat[]{\centering\includegraphics[scale=0.69]{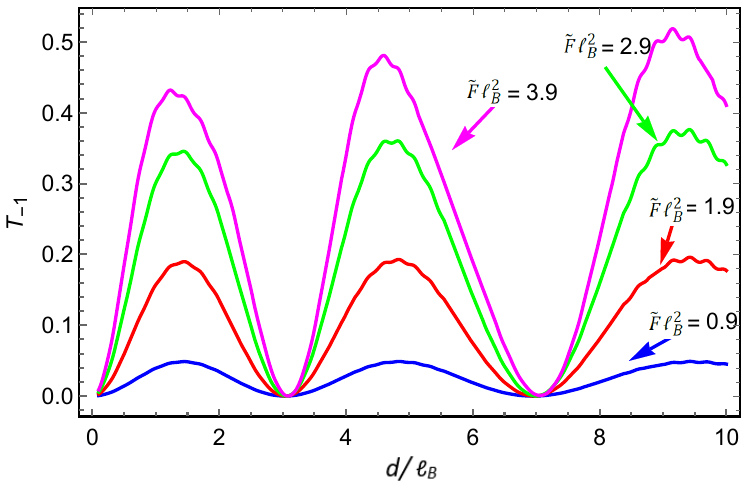}\label{fig5d}}
\caption{{(Color online) Transmission probabilities as a function of the barrier width $d/\ell_B$, with $\varepsilon \ell_B=15$ and $\Delta \ell_B=0$, (a) $\varpi \ell_B=2$, $k_y \ell_B=1$, $\tilde{F} \ell_B^2=2.9$, (b):  $\varpi \ell_B=2$, $\tilde{F} \ell_B^2=2.9$ and $k_y\ell_B=\{0, 1, 2, 3\}$, (c): $k_y \ell_B=1$, $\tilde{F} \ell_B^2=2.9$ and $\varpi \ell_B=\{1.8, 2, 2.2, 2.4\}$, (d): $k_y \ell_B=1$, $\varpi \ell_B=2$ and $\tilde{F} \ell_B^2=\{0.9, 1.9, 2.9, 3.9\}$.}}\label{fig5}
\end{figure}
Fig. \ref{fig5} shows the transmission probabilities as a function of the barrier width $d/\ell_B$. {In} Fig. \ref{fig5a} we observe that all the transmissions have sinusoidal behavior. The total transmission oscillates in the vicinity of {one} (Klein paradox). $T_0$ is predominant and its {oscillation} amplitude decreases when the width increases. The transmissions with photon exchange also oscillate, but with phase shift, which increases along the $d/\ell_B$-axis. For certain values of $d/\ell_B$, the transmissions with or without photon exchange are equal.
Fig. \ref{fig5b} displays transmission with photon emission for different values of the transverse wave vector $k_y\ell_B$. There is always a sinusoidal behavior with increasing amplitude along the $d/\ell_B$-axis. When $k_y\ell_B$ increases, the width of the oscillations {decreases}. 
In Fig. \ref{fig5c}, we show the effect of the laser field frequency on transmission. {We} notice that the  amplitude and period of oscillations decrease as the frequency increases. {Thus}, the increase in frequency suppresses the transmissions with photon exchanges.
We vary the intensity of the laser field $\tilde{F} \ell_B^2$ in Fig. \ref{fig5d} and observe that the transmission is oscillating with the same period. We notice that the increase in $\tilde{F} \ell_B^2$ causes an increase in transmission  with photon exchange and   decreases that  of the central band.

\begin{figure}[ht]
	\centering	
	\subfloat[]{\centering\includegraphics[scale=0.66]{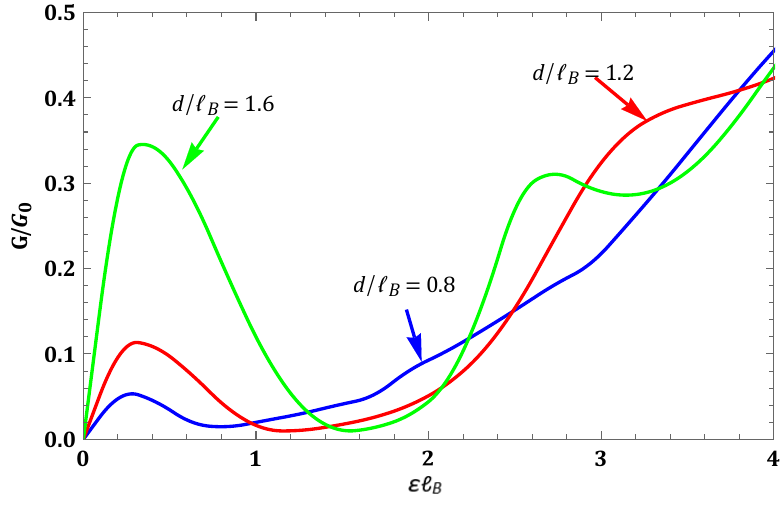}	\label{fig6a}} 	
	\subfloat[]{\centering\includegraphics[scale=0.66]{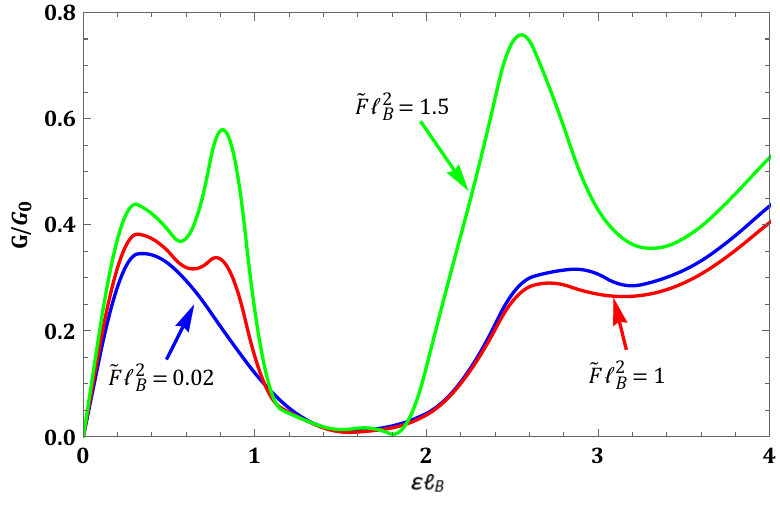}\label{fig6b}}\\		\subfloat[]{\centering\includegraphics[scale=0.66]{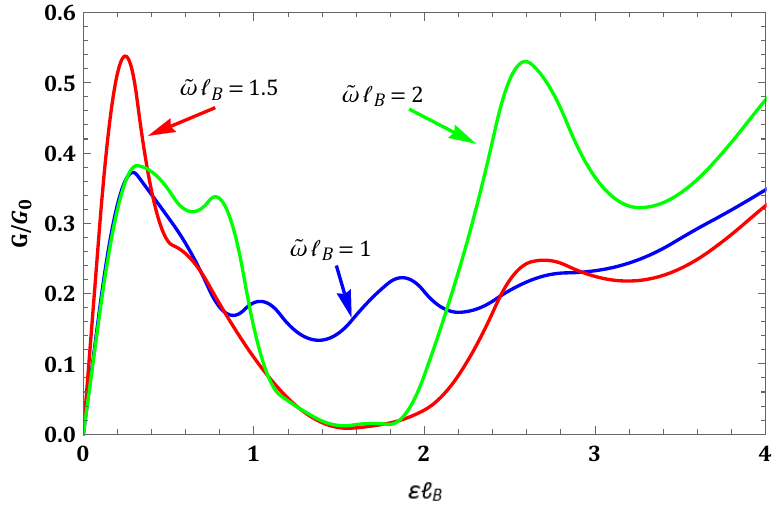}	\label{fig6c}} 
	\subfloat[]{\centering\includegraphics[scale=0.66]{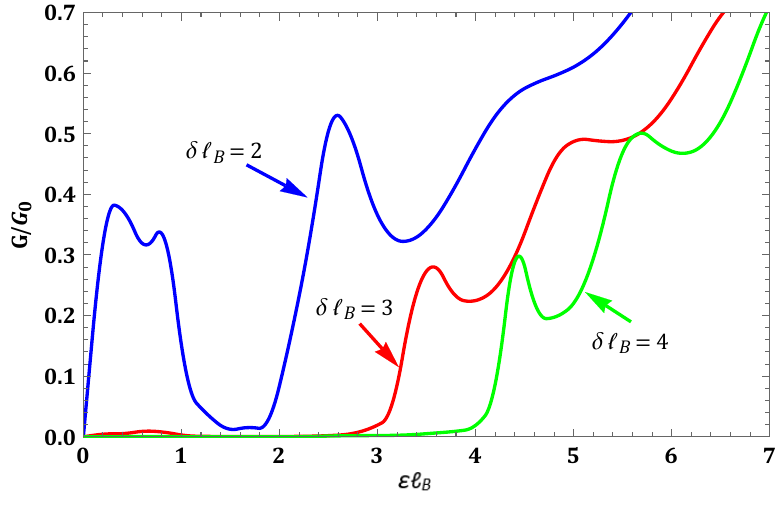}\label{fig6d}}
	\caption{{(Color online)  Zero temperature conductance as a function of $\varepsilon \ell_B$, (a) $\varpi \ell_B=1$, $\tilde{F} \ell_B^2=0.03$, $\delta \ell_B=2$ and $\frac{d}{\ell_B}=\{0.8, 1.2, 1.6\}$, (b): $\varpi \ell_B=2$, $\frac{d}{\ell_B}=1.6$, $\delta \ell_B=2$ and $\tilde{F} \ell_B^2=\{0.02, 1, 1.5\}$, (c): $\tilde{F} \ell_B^2=1$, $\frac{d}{\ell_B}=1.6$, $\delta \ell_B=2$ and $\varpi \ell_B=\{1, 1.5, 2\}$, (d): $\varpi \ell_B=2$, $\frac{d}{\ell_B}=1.6$, $\tilde{F} \ell_B^2=1$ and $\delta \ell_B=\{2, 3, 4\}$.}}\label{fig6}
\end{figure}

\newpage
 In Fig. \ref{fig6}, we plot the conductance  as a function of the energy  $\varepsilon \ell_B$. {Choosing} different values of width $\frac{d}{\ell_B}$, Fig. \ref{fig6a}  reveals that the conductance varies almost exponentially for lower values of $\frac{d}{\ell_B }$, and oscillates  when $\frac{d}{\ell_B}$ increases.
 Fig. \ref{fig6b} shows  the effect of  intensity $\tilde{F}\ell_B^2$ of the laser field  on conductance. We observe  that conductance increases as $\tilde{F}\ell_B^2$ increases, but it vanishes when $\varepsilon \to\delta $. 
 Fig. \ref{fig6c} is plotted for different values of frequency $\varpi \ell_B$.  We notice that the conductance tends to zero when $\varepsilon \ell_B$ is close to $\delta \ell_B$ and the oscillations increase as $\varpi  \ell_B$ increases. 
 In Fig. \ref{fig6d}, we vary $\delta \ell_B$ to observe  that the conductance is always almost zero when $\varepsilon$ tends towards $\delta$.
 {\bb Ultimately, to enhance conductance, it is essential to augment the quantity of electrons traversing the barrier, consequently elevating transmission. As observed, transmission rises with an increase in incident energy or a reduction in barrier width, and it also climbs when the intensity of the laser field decreases or its frequency rises.}
 
%  Finally, to increase the conductance, it is necessary to increase the number of electrons crossing the barrier, thereby increasing the transmission. As we have seen, the transmission increases when the incident energy increases or the barrier width decreases, as well as when the intensity of the laser field decreases or its frequency increases.

% as we saw above for the transmission probability because the conductance is an integral of the total transmission.

\begin{figure}[ht]
	\centering	
	\subfloat[]{\centering\includegraphics[scale=0.62]{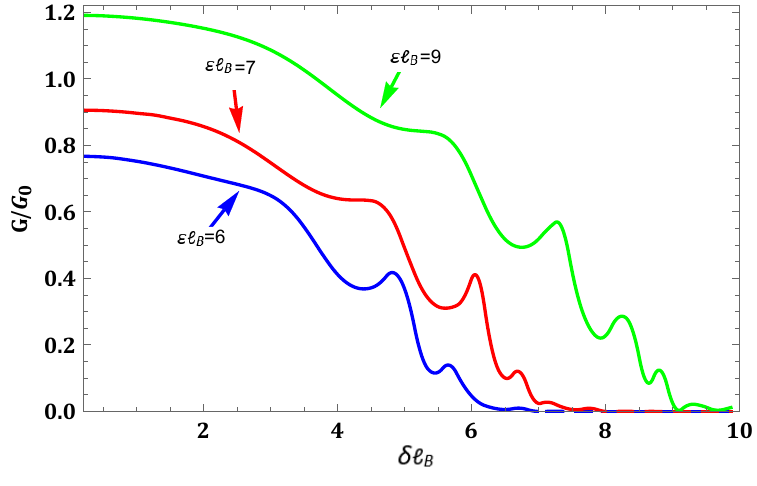}\label{fig7a}}  \ \ \ 
	\subfloat[]{\centering\includegraphics[scale=0.62]{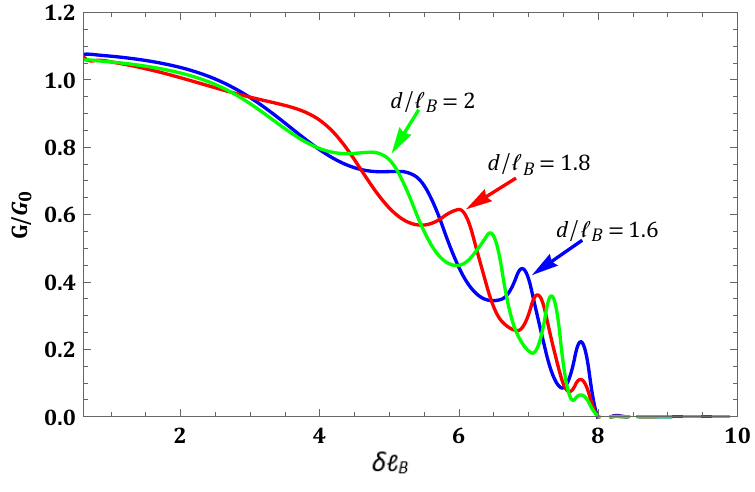}\label{fig7b}}\\
	\subfloat[]{\centering\includegraphics[scale=0.62]{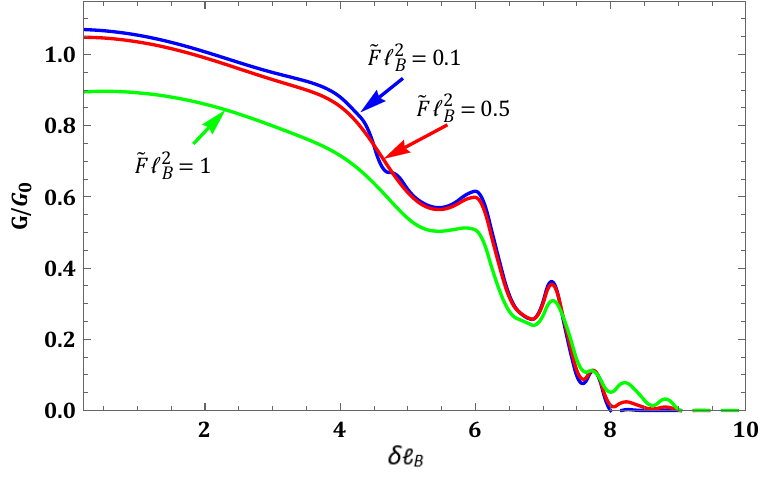}\label{fig7c}}  \ \ \ 
	\subfloat[]{\centering\includegraphics[scale=0.62]{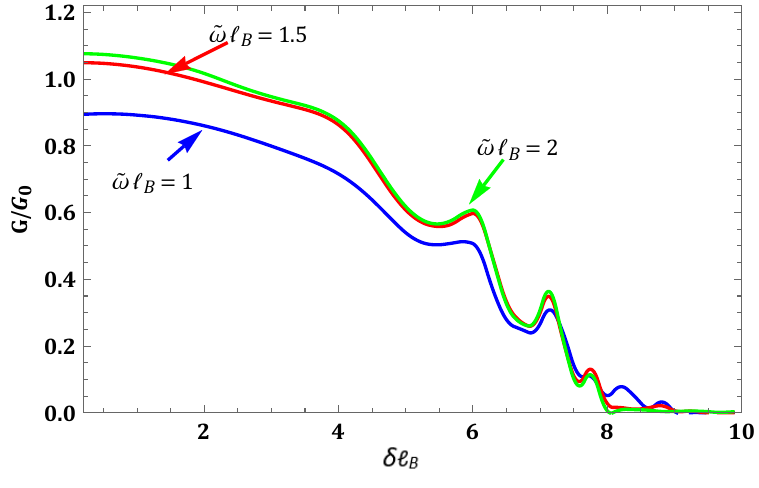}\label{fig7d}}
	\caption{{(Color online)  Zero temperature conductance as a function of $\delta \ell_B$, (a) $\varpi \ell_B=1$, $\tilde{F} \ell_B^2=0.5$, $\frac{d}{\ell_B}=1.8$ and $\varepsilon \ell_B=\{6, 7, 9\}$, (b): $\varpi \ell_B=1$, $\tilde{F} \ell_B^2=0.03$, $\varepsilon \ell_B=8$ and $\frac{d}{\ell_B}=\{1.6, 1.8, 2\}$, (c): $\varpi \ell_B=1$, $\frac{d}{\ell_B}=1.8$, $\varepsilon \ell_B=8$ and $\tilde{F} \ell_B^2=\{0.1,0.5,1\}$, (d): $\tilde{F} \ell_B^2=1$, $\frac{d}{\ell_B}=1.8$, $\varepsilon \ell_B=8$ and $\varpi \ell_B=\{1, 1.5, 2\}$.}}\label{fig7}
\end{figure}

In Figure \ref{fig7}, the conductance is represented as a function of the energy gap $\delta \ell_B$. By choosing three values of incident energy in Fig. \ref{fig7a}, we show  that the conductance is maximum at the beginning, then decreases in an oscillatory way towards zero near the value $\delta =\varepsilon $. The amplitude increases when incident energy increases as well, exhibiting a {behavior similar} to transmission as we have seen before. 
Fig. \ref{fig7b} shows the effect of width $\frac{d}{\ell_B}$ on the conductance. There are always resonance peaks that appear around $\delta \ell_B=3$, the number of oscillations increases with the increase of $\frac{d}{\ell_B}$. In  Figs. \ref{fig7c} and \ref{fig7d}, we visualize the effect of the laser field parameters on the conductance. They show that the amplitude of the conductance increases with the increase in frequency, and decreases when the amplitude increases.

\newpage
\section{Conclusion}

We studied the effect of a gapped magnetic barrier irradiated by a laser field generated by an electric field of amplitude $F$ and frequency $\omega$ on Dirac fermions in graphene. We started with the solution of the eigenvalue equations to determine the spinors in the three regions {of the gapped sheet}. We used the Floquet theory, and the solution of Weber's differential equation to determine the eigenspinors corresponding to each region as combinations of parabolic cylindrical functions. Then we employed the boundary conditions, which give four equations, each equation has infinite modes. To solve them, we used the transfer matrix approach to obtain a matrix of infinite order that is difficult to solve. For simplicity, we focused only on the three first bands, the central band corresponds to $l=0$ and the two first side bands correspond to $l=\pm1$. Lastly, we calculated the integral of the total transmission probability to obtain conduction at zero temperature.

When a barrier oscillates in time, it generates several energy bands, namely the photon exchange between the barrier and the Dirac fermions. Here we found that the transmission process with zero photon exchange is much more important than the process with photon exchange. Klein's paradox is still present, but we can suppress it. As we know, the original Klein effect is only observed for normal incidences ($\phi_0=0$), but in this work, this effect is observed for non-normal incidences. When the barrier width is increased, the transmission decreases until it disappears for a critical width, the same thing happens for the conductance. On the other hand, the transmission increases when the incident energy increases. However, to have transmission, it is necessary to satisfy the condition that binds the incident energy to the other barrier parameters: $\varepsilon >\frac{\frac{d}{\ell_B^2}-l \varpi}{1+\sin{\phi_0}}$. {As we know the conductance exists if we have a non-zero transmission, which always implies the verification of this last condition.}.

\end{document}